\documentclass{PoS}

\usepackage{gensymb}
\usepackage{graphics}

\newcommand{\gevc}{\mathrm{GeV}/c}
\newcommand{\gev}{\mathrm{GeV}}
\newcommand{\mevc}{\mathrm{MeV}/c}

\newcommand{\pt}{p_{\rm T}}
\newcommand{\DtoKpi}{{\rm D^0\to K^-\pi^+}}
\newcommand{\DtoKpipi}{{\rm D^+\to K^-\pi^+\pi^+}}
\newcommand{\DstartoDpi}{{\rm D^{*+}\to D^0\pi^+}}
\newcommand{\Dzero}{{\rm D^0}}
\newcommand{\Dstar}{{\rm D^{*+}}}
\newcommand{\Dplus}{{\rm D^+}}
\newcommand{\Ds}{{\rm D_s^+}}
\newcommand{\Dstophipi}{{\rm D_s^+\to \phi\pi^+}}

\newcommand {\pT}        {\ensuremath{p_{\mathrm{\textsc{t}}}}}
\newcommand {\sqrtSnn}   {\ensuremath{\sqrt{s_{\textsc{nn}}}}}
\newcommand {\sqrtS}     {\ensuremath{\sqrt{s}}}

\newcommand {\dpt}       {\ensuremath{\mathrm{d}\pT }}

\newcommand {\Rppb}       {\ensuremath{R_\mathrm{pPb}}}
\newcommand {\Qppb}       {\ensuremath{Q_\mathrm{pPb}}}
\newcommand {\dEdx}      {\ensuremath{\mathrm{d}E/\mathrm{d}x }}

\newcommand {\mass}     {\mbox{\rm MeV$\kern-0.15em /\kern-0.12em c^2$}}
\newcommand {\tev}      {\mbox{${\rm TeV}$}}

\newcommand{\Jpsi}      {\mbox{J\kern-0.02em /\kern-0.05em$\psi$}}
\newcommand{\ccbar}     {\mbox{$\mathrm {c\overline{c}}$}}

\title{Open heavy-flavour measurements in pp and p--Pb collisions with ALICE at the LHC}

\ShortTitle{Open heavy-flavour measurements in pp and p--Pb collisions with ALICE at the LHC}

\author{\speaker{Jaime Norman} for the ALICE collaboration\\
	\\
	The Oliver Lodge Laboratory, The University of Liverpool, Liverpool L69 7ZE, United Kingdom\\
	E-mail: \email{jaime.norman@cern.ch}}

\abstract{The ALICE detector is well suited to measure heavy-flavour (charm and beauty) production via hadronic and semi-leptonic decay channels of heavy-flavour particles. Here an overview of heavy-flavour measurements made with the ALICE detector during Run 1 in pp and p--Pb collisions is presented and discussed.}

\FullConference{55th International Winter Meeting on Nuclear Physics\\
	23-27 January, 2017\\
	Bormio, Italy}

\begin{document}
	
	\section{Introduction}
	
	Heavy quarks (charm, beauty) are produced in hard partonic scattering processes in high energy nucleon--nucleon collisions, with large $Q^2$ which, along with their large mass, means their production is calculable using perturbative Quantum Chromodynamics (pQCD) down to low transverse momentum. 
	The production of a given final state hadron $H$ is calculated assuming long-range and short-range effects can be factorised into the initial parton distribution of the colliding hadrons $f_a(x_1,Q^2)$ and $f_a(x_1,Q^2)$, the partonic production cross section $\sigma_{ab\rightarrow q\bar{q}}$ and the fragmentation function of a quark to a hadron $D_{q\rightarrow H}(z_q,Q^2)$, such that
	
	\begin{equation}
	\sigma_{hh\rightarrow H} = f_a(x_1,Q^2) \otimes f_b(x_2,Q^2) \otimes \sigma_{ab\rightarrow q\bar{q}} \otimes D_{q\rightarrow H}(z_q,Q^2)
	\end{equation}
	
	The measurement of heavy-flavour production in pp collisions thus acts as an important testing ground for state-of-the-art pQCD calculations.
	Measurements in pp collisions also offer an essential reference for measurements in Pb--Pb collisions, where charm and beauty quarks are expected to lose energy due to the hot matter known as the Quark--Gluon Plasma (QGP) created in these collisions, where quarks are deconfined.
	
	Heavy-flavour production in p--Pb collisions offers a way to study so called `cold nuclear matter effects', which include modifications to the parton distribution in the nucleus, and possible final state effects such as multiple scattering leading to energy loss at low transverse momentum. These measurements also provide a way to separate cold nuclear matter effects in Pb--Pb collisions from the `hot' effects due to the deconfined medium created in these collisions.
	
	The measurement of heavy-flavour production as a function of event multiplicity can also give important insight into the correlation and interplay between the `hard' part of the event (hard partonic scattering leading to heavy-flavour production), and the `soft' part (softer `minijets' from multiple parton interactions, initial- and final-state radiation and/or fragmentation of beam remnants, all contributing to the final event multiplicity).
	The measurement of the azimuthal correlation between heavy-flavour particles and charged particles can offer insight into charm production and fragmentation processes.
	These measurements in pp and p--Pb collisions offer a way to further constrain descriptions of heavy-flavour production.
	In addition, recent observation of the collective behaviour of charged particles in p--Pb collisions and high-multiplicity pp collisions means the measurement of heavy-flavour production is also a good testing ground to understand whether heavy quarks also build up collectivity in smaller systems, and to further understand the physical mechanisms behind the collective behaviour.
	
	The ALICE detector \cite{Aamodt:2008zz} offers excellent tracking and PID capabilities, allowing for the measurement of heavy-flavour production in a variety of channels over a wide transverse momentum range, including the full reconstruction of D mesons in the decay channels $\DtoKpi$, $\DtoKpipi$, $\DstartoDpi$ and $\Dstophipi$ at mid-rapidity, the measurement of electrons from heavy-flavour decays (measuring inclusive electrons from charm and beauty, or separating the beauty component) at mid-rapidity, and the measurement of muons from heavy-flavour decays at forward-rapidity. Here we present a selection of recent heavy-flavour measurements in pp and p--Pb collisions from ALICE. In section \ref{sec:experiment} an overview of the ALICE apparatus is presented. In section \ref{sec:results} heavy-flavour results are shown. Finally section \ref{sec:summary} will give a summary and an outlook.
	
	\section{Experimental apparatus}
	\label{sec:experiment}
	
	The key features of the ALICE detector include high-resolution tracking performance down to very low momentum ($100~\mevc$) and excellent Particle Identification (PID) capabilities over a wide momentum range opimised for a high-multiplicity environment. The central barrel of ALICE features the Inner Tracking System (ITS), the Time Projection Chamber (TPC) and the Time-of-Flight detector (TOF) covering the entire azimuthal range, and designed to track and identify charged particles. The ITS is a six layer silicon detector close to the beampipe.
	The TPC is equiped with multiwire proportional chambers, which allow for the tracking of particles and the measurement of the specific energy loss $\dEdx$ of particles for charged-particle PID.
	The TOF detector is based on Multigap Resistive Plate Chamber technology and performs PID via the measurement of charged particle time-of-flight. Electrons can additionally be identified by the Electromagnetic Calorimeter (EMCal) and the Transition Radiation Detector (TRD).
	
	The forward muon arm is located at forward angles ($2\degree$ -- $9\degree, 4 < \eta < 2.5$) and consists of absorbers to filter muons, ten layers of high-granularity, cathode strip tracking stations, and four layers of resistive plate chambers used for muon triggering. The muon arm is placed inside a $3$ Tm dipole magnet.
	
	D mesons are measured through an invariant mass analysis after PID and selection on their decay topologies exploiting the decay vertex displacement from the primary vertex, and those from B-hadron decays are subtracted using pQCD predictions \cite{Cacciari:1998it} of charm and beauty production. Electrons from heavy-flavour decays are measured by subtracting background contributions 
	from the inclusive electron spectra, and electrons from beauty-decays can also be separated via their impact parameter distribution. Muons from heavy-flavour decays are measured subtracting muons from light-flavour and W decays.
	
	\section{Results}
	\label{sec:results}
	
\begin{figure}[!hbt]
		\centering
		\includegraphics[width=.45\textwidth]{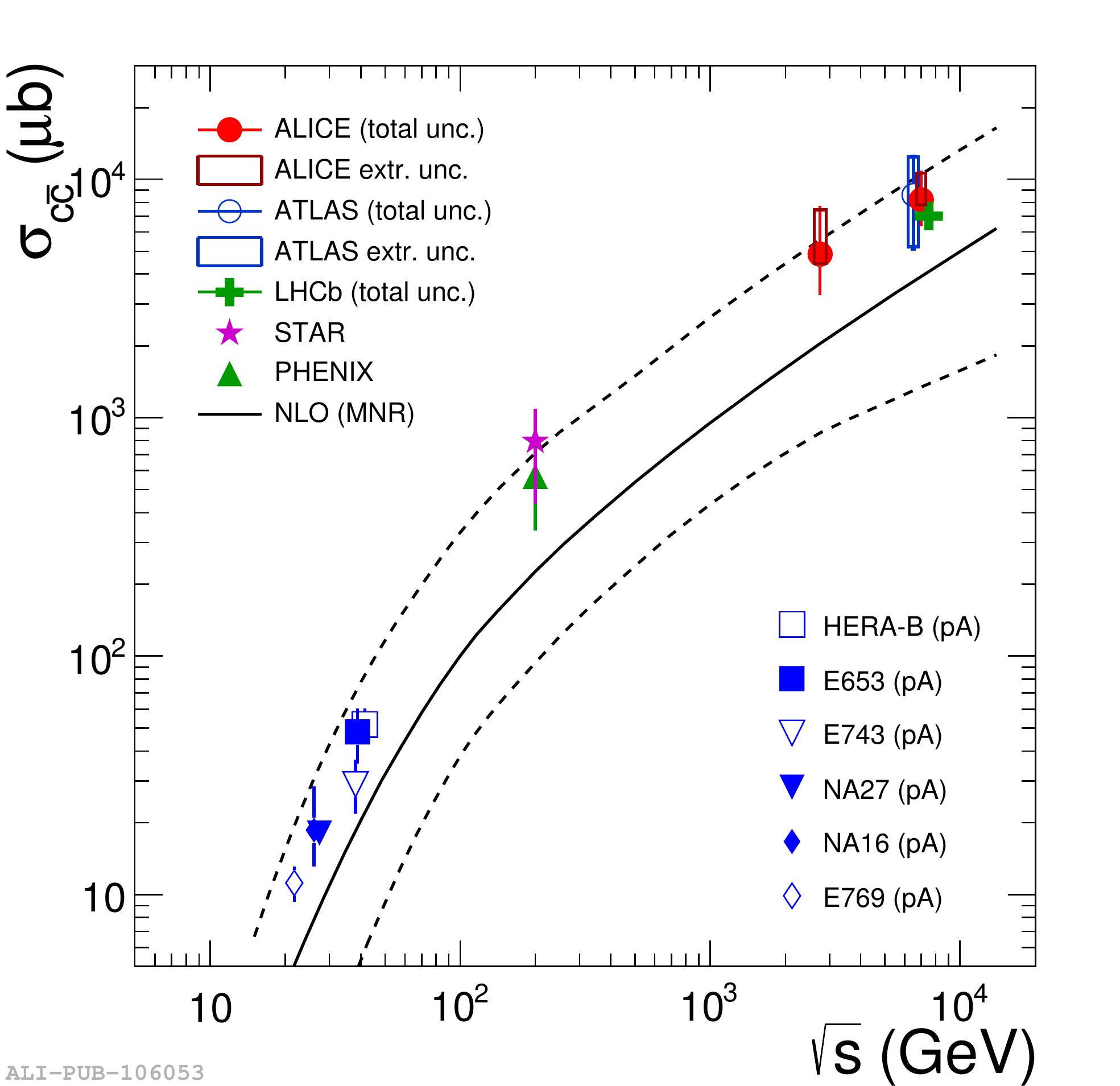}
		\caption[$\bar{c}c$ production cross section in nucleon--nucleon collisions as a function of $\sqrtS$]{Total inclusive charm production cross section in nucleon--nucleon collisions
			as a function of
			$\sqrtS$ \cite{Adam:2016ich}. Data are from pA collisions for $\sqrt s<100$~GeV and from pp collisions for $\sqrt s>100~\gev$. Data from pA collisions were scaled by $1/A$. Results from NLO pQCD calculations (MNR \cite{MANGANO1992295}) and their uncertainties are shown as solid and dashed lines.}  
		\label{fig:total-charm}
	\end{figure}

	The $\pt$-differential cross section of $\Dzero$, $\Dplus$, $\Dstar$ and $\Ds$ mesons \cite{Adam:2016ich}, muons from heavy-flavour decays \cite{Abelev:2012qh,Abelev:2012pi}, electrons from heavy-flavour decays \cite{Abelev:2012xe,Abelev:2014gla} and beauty-decay electrons \cite{Abelev:2012sca,Abelev:2014hla} have been measured in pp collisions at $\sqrtS = 2.76~\tev$ and $\sqrtS = 7~\tev$ and are found to be compatible with pQCD predictions including calculations from FONLL \cite{Cacciari:1998it}, GM-VFNS \cite{Kniehl:2005mk,Kniehl:2012ti} and the $k_T$-factorisation scheme \cite{Maciula:2013wg}.
	The $\Dzero$ production cross section has also been measured down to $\pt = 0$ at $\sqrtS = 7~\tev$ \cite{Adam:2016ich}. 
	The total $\ccbar$ cross section can be derived from these measurements, and the 
	scaling of the total inclusive charm production cross section in pp and p--A collisions as a function of collision energy $\sqrtS$ is shown in figure \ref{fig:total-charm}. The total cross section is described within uncertainties by NLO pQCD calculations \cite{MANGANO1992295} with the measurements falling on the upper edge of the theoretical uncertainty.
	
\begin{figure}[!hbt]
		\centering
		\includegraphics[width=.43\textwidth]{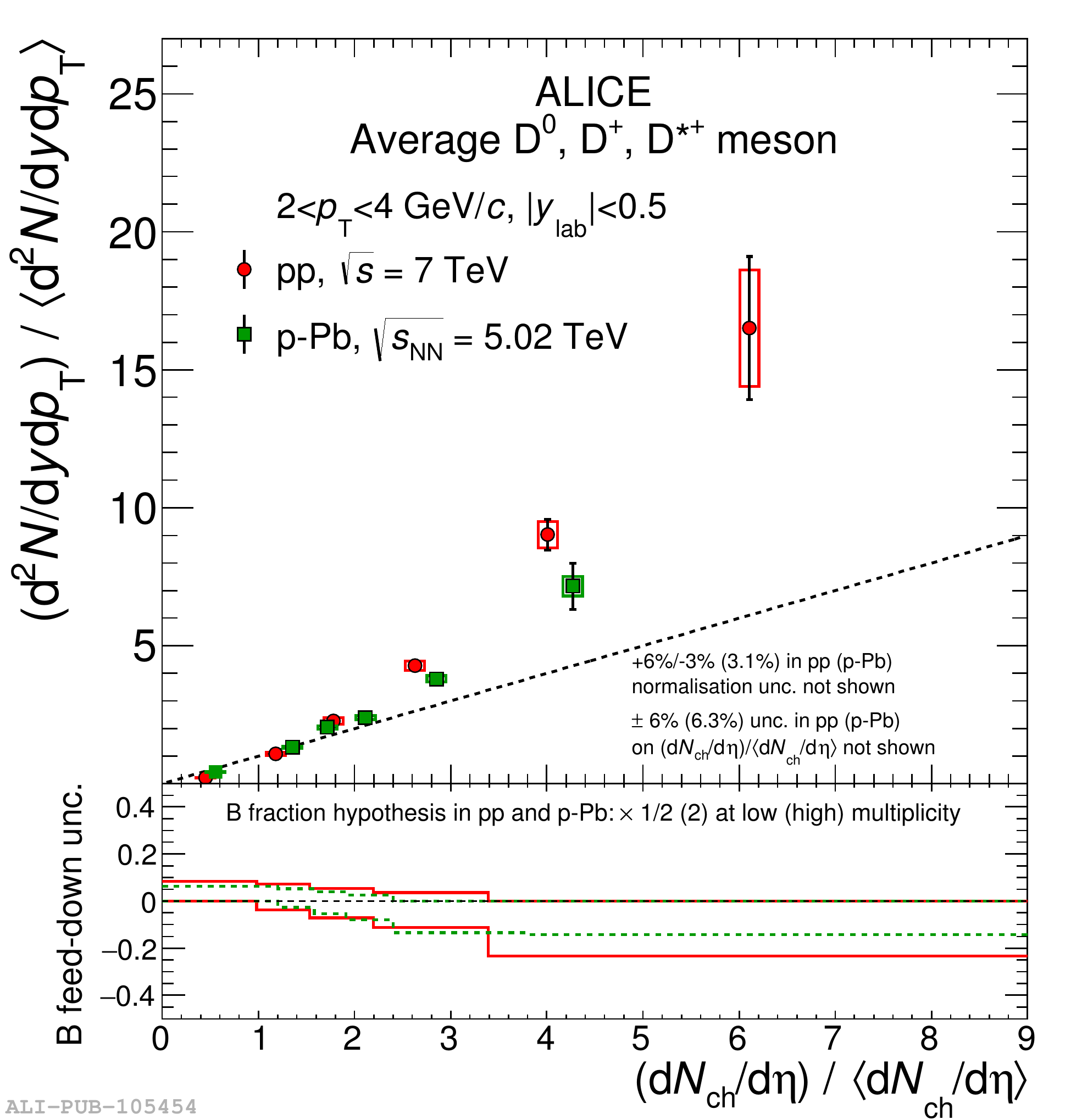}
		\includegraphics[width=.43\textwidth]{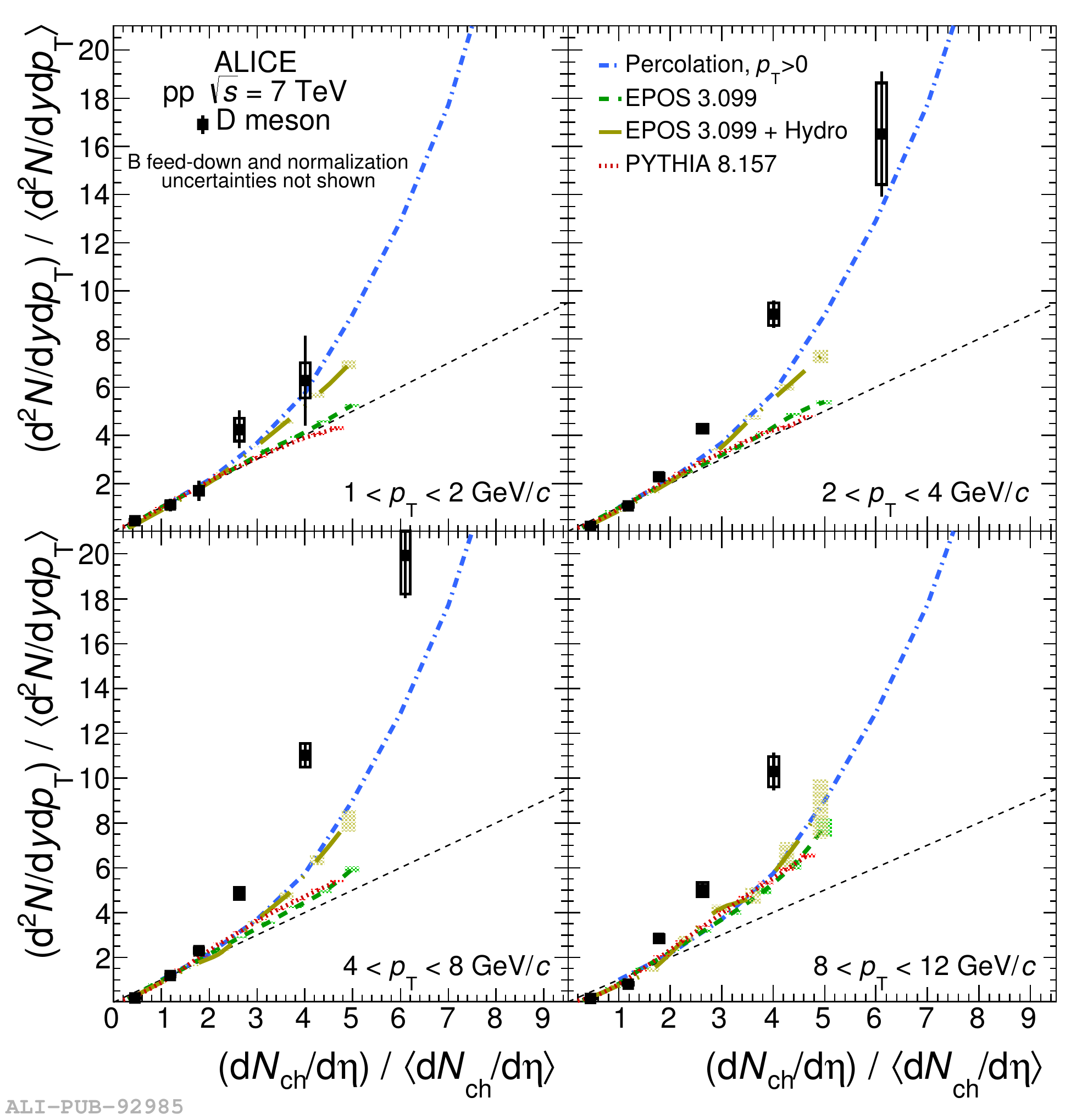}
		\caption[D meson yield vs. multiplicity]{Left: Average self-normalised D meson yields as a function of the self-normalised charged particle multiplicity at mid-rapidity for $2 < \pt < 4~\gevc$ in pp and p--Pb collisions \cite{Adam:2016mkz}. The uncertainty on the B feed-down fraction is drawn on the bottom panel. Right: The measurement in pp collisions is compared to models \cite{Adam:2015ota}.}  
		\label{fig:Dmeson-yield-vs-multiplicity}
	\end{figure}

	Heavy-flavour production has been studied as a function of the event multiplicity (measured at mid-rapidity) via the measurement of the self normalised yields - the corrected per-event yield in a given multiplicity interval normalised to the multiplicity-integrated per-event yield.
	Figure \ref{fig:Dmeson-yield-vs-multiplicity} shows the average self-normalised D meson yield as a function of the self-normalised charged particle multiplicity, for pp collisions at $\sqrtS = 7 ~\tev$ \cite{Adam:2015ota} and p--Pb collisions at $\sqrtSnn = 5.02~\tev$ \cite{Adam:2016mkz}. A faster than linear increase is seen for both collision systems, which is compared with models for pp collisions on the right of figure \ref{fig:Dmeson-yield-vs-multiplicity}. The data qualitatively agree with models including parton saturation and interaction between colour strings in the percolation model \cite{Ferreiro:2012fb}, and multi-parton interactions via the PYTHIA8 \cite{Sjostrand:2007gs} and EPOS \cite{Werner:2013tya} event generators. The measurement in p--Pb collisions has also been compared with predictions from EPOS which can be found in \cite{Adam:2016mkz}. It is interesting to note that switching viscous hydrodynamic effects on for EPOS gives a better description of the steepness of the measured data in both pp and p--Pb collisions, suggesting collective effects could play a role in particle production at high multiplicity.

\begin{figure}[!hbt]
		\centering
		\includegraphics[width=.40\textwidth]{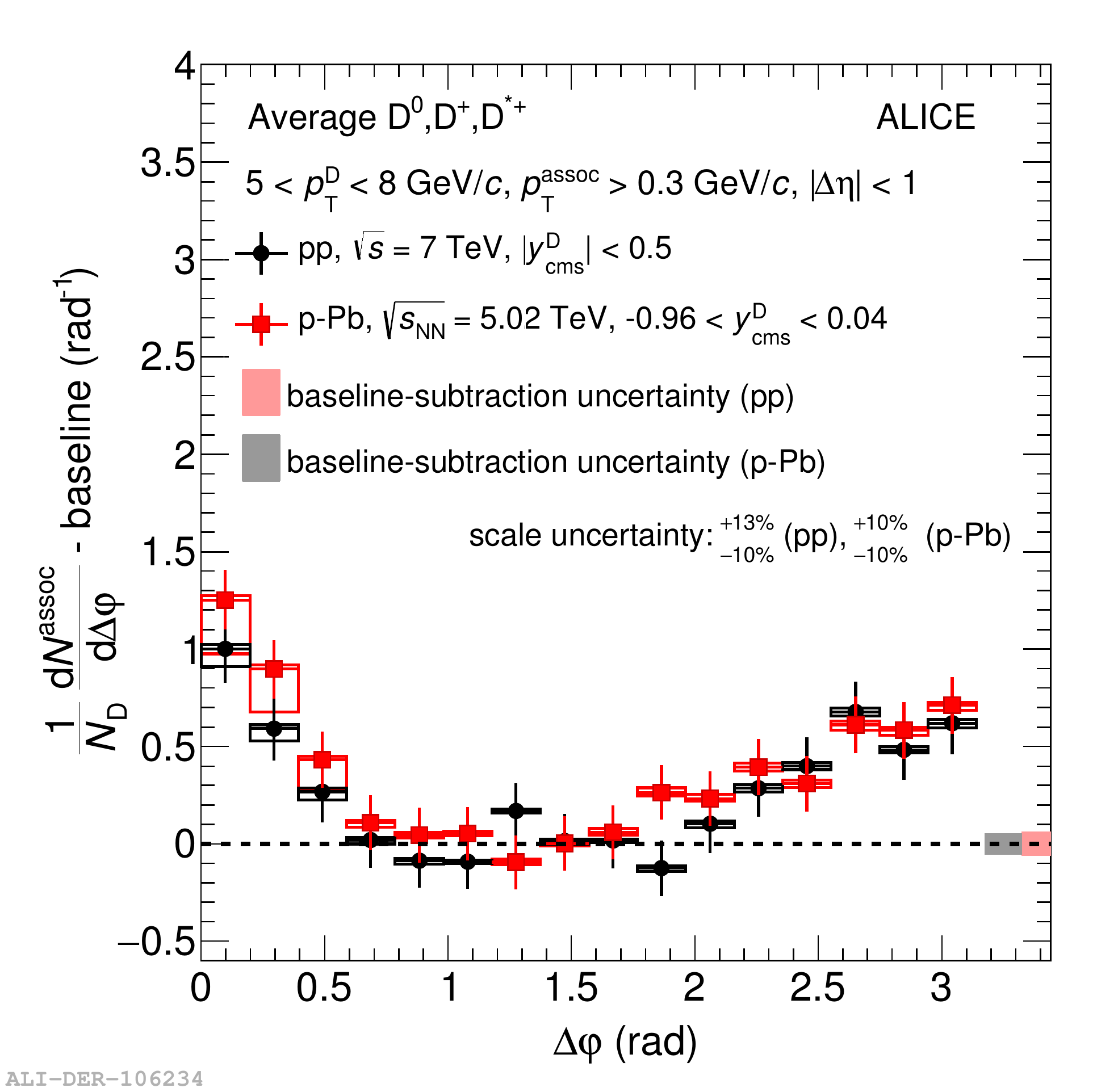}
		\includegraphics[width=.45\textwidth]{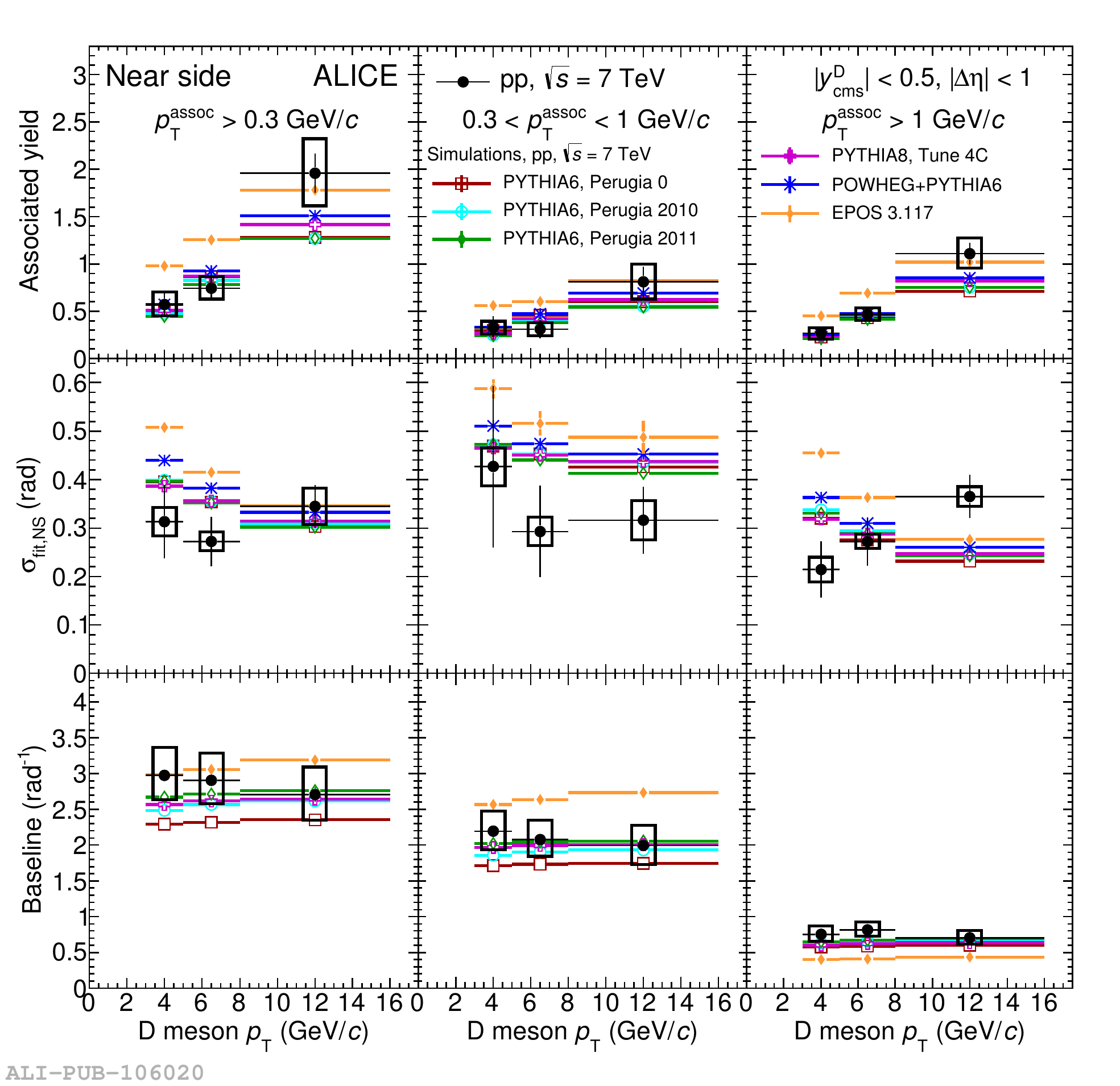}
		\caption[Azimuthal correlation of D mesons with charged particles]{Left: Comparison of the azimuthal correlation distributions of D mesons with $5 < p^{\mathrm{D}}_{\mathrm{T}} < 8~\gevc$ and charged particles with $p^{\mathrm{assoc}}_{\mathrm{T}} > 0.3~\gevc$ in pp collisions at $\sqrtS = 7~\tev$ and p--Pb collisions at $\sqrtSnn = 5.02~\tev$ \cite{ALICE:2016clc}. Right: Comparison of the near-side peak associated yield (top), the near-side peak width (middle) and the baseline (bottom) measured in pp collisions at $\sqrtS = 7~\tev$ with the fit parameters obtained from different Monte-Carlo event generators
		\cite{Sjostrand:2006za,Sjostrand:2007gs,Sjostrand:2007gs,Nason:2004rx,Frixione:2007vw}.}  
		\label{fig:azimuthal-Dch}
	\end{figure}

	The angular correlation of D mesons with charged particles (i.e. the distribution of differences in azimuthal angles $\Delta \phi = \phi_\mathrm{ch} - \phi_\mathrm{D}$ and pseudorapidities $\Delta \eta = \eta_\mathrm{ch} - \eta_\mathrm{D}$ ) has been measured by ALICE in pp collisions at $\sqrtS = 7 ~\tev$ and in p--Pb collisions at $\sqrtSnn = 5.02 ~\tev$ \cite{ALICE:2016clc}.
	Figure \ref{fig:azimuthal-Dch} (left) shows the averaged $\Dzero$, $\Dplus$ and $\Dstar$ correlation distribution in both collision systems. The distribution shows a nearside peak at $\Delta\phi=0$ and a broader away side peak at $\Delta\phi=\pi$, and the pp and p--Pb distributions agree within uncertainties.
	Figure \ref{fig:azimuthal-Dch} (right) shows parameters associated to the fit to the azimuthal correlation distribution including the associated yield and width of the nearside peak and the baseline of the distribution in pp collisions, where it is seen that the fit to data agrees with the same fit parameters obtained from Monte-Carlo generators including PYTHIA6 \cite{Sjostrand:2006za} with different tunes, PYTHIA8 \cite{Sjostrand:2007gs}, POWHEG \cite{Nason:2004rx,Frixione:2007vw} coupled to PYTHIA, and EPOS \cite{Werner:2013tya}. 
	Collectivity can be quantified by the second Fourier coefficient of the azimuthal asymmetry $v_\mathrm{2}$, and if the charm quark participated in the collective expansion of the system, a $v_\mathrm{2}$-like modulation would introduce a bias in the near side yield and width of the nearside peak.
	The extreme assumption of $v_\mathrm{2} = 0.05$ for D mesons and $v_\mathrm{2} = 0.05 (0.1)$ for associated charged particles with $\pt > 0.3 (1)~\gevc$ would modify the associated near-side yield by $\sim 10\%$ and the width of the near-side peak by $<4\%$ for D mesons with $5 < p^{\mathrm{D}}_{\mathrm{T}} < 8~\gevc$ and for associated charged particles with $0.3 < p^{\mathrm{assoc}}_{\mathrm{T}} < 1~\gevc$, which is outside the precision of this measurement, but this will be interesting to study in future measurements with higher statistics datasets.
	
\begin{figure}[hbt]
		\centering
		\includegraphics[width=.42\textwidth]{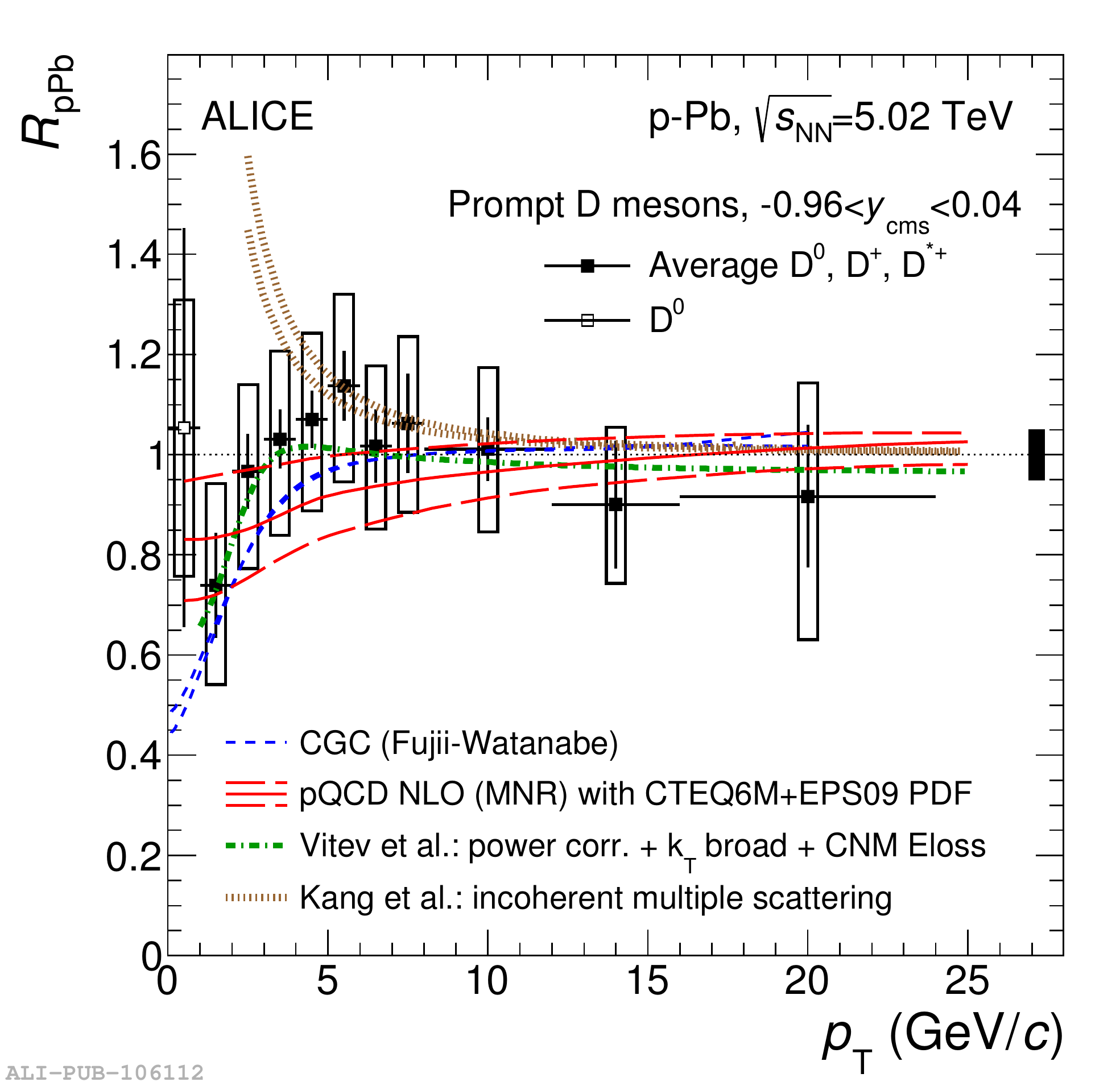}
		\includegraphics[width=.42\textwidth]{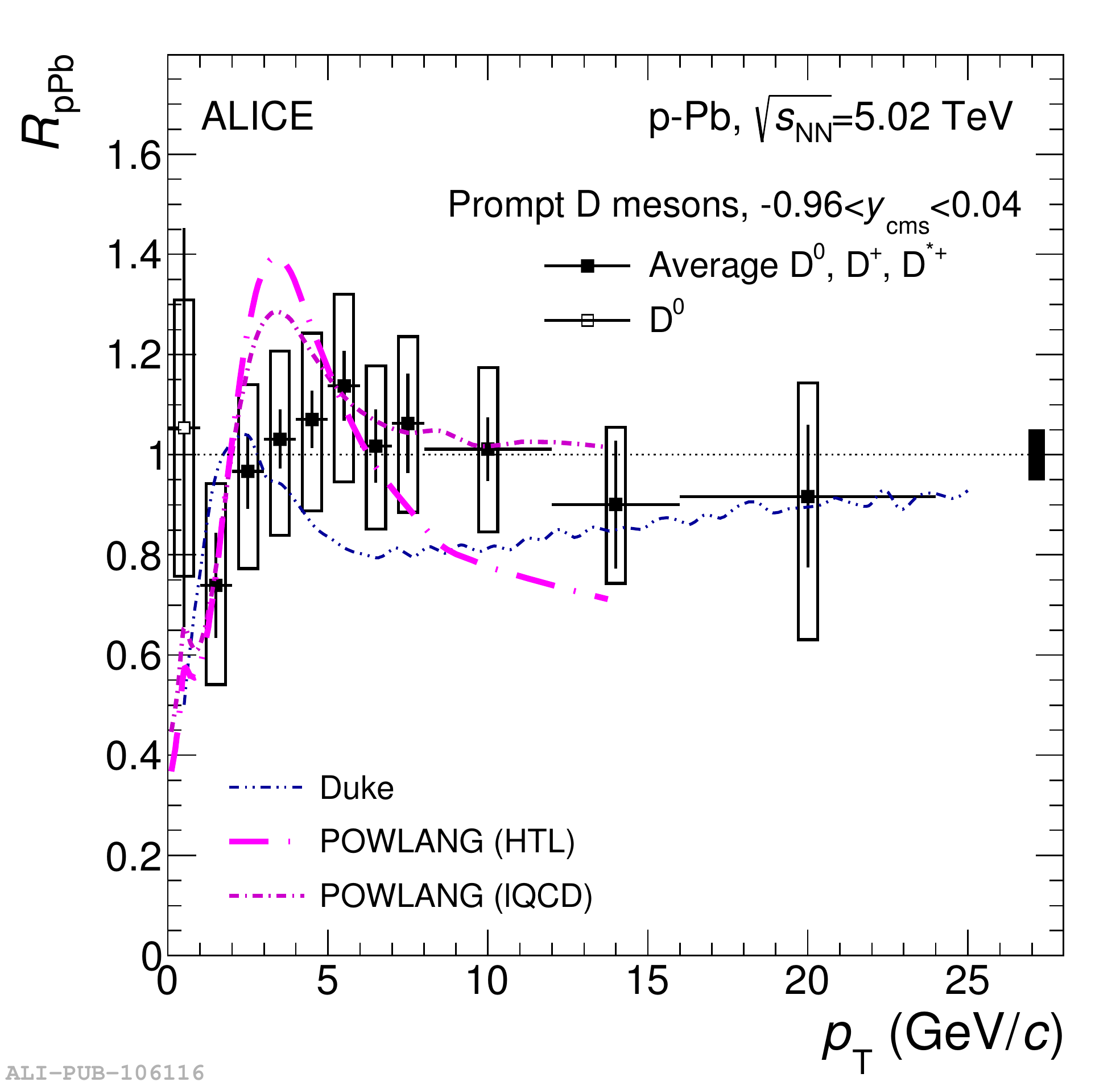}
		\caption[$\Rppb$ of D mesons measured by ALICE]{The averaged D-meson nuclear modification factor $\Rppb$ \cite{Adam:2016ich} compared with models that include cold nuclear matter effects \cite{MANGANO1992295,Eskola:2009uj,Sharma:2009hn,Fujii:2013yja,Kang:2014hha} (left) and models that include a small QGP formation \cite{Xu:2015iha,Beraudo:2015wsd} (right).}  
		\label{fig:D0RpPb-ALICE}
	\end{figure}

	The modification of the D-meson yield in p--Pb collisions with respect to pp collisions is quantified by the nuclear modification factor $\Rppb$, which in minimum-bias collisions is measured as a function of $\pt$ as
	
	\begin{equation}
	\Rppb = \frac{1}{A} 
	\frac{\mathrm{d}\sigma_\mathrm{pPb} / \mathrm{d}\pt}
	{\mathrm{d}\sigma_\mathrm{pp} / \mathrm{d}\pt}
	\end{equation}
	
	where $\mathrm{d}\sigma_\mathrm{pPb(pp)} / \mathrm{d}\pt$ is the $\pt$-differential cross section of a given D-meson species in \mbox{p--Pb}(pp) collisions and $A = 208$ is the mass number of lead. The nuclear modification factors $\Rppb$ of D mesons (including $\Dzero, \Dplus$, $\Dstar$ and $\Ds$) have been measured at ALICE at mid-rapidity \cite{Adam:2016ich}, and are found to be compatible with each other. Figure \ref{fig:D0RpPb-ALICE} shows the averaged $\Dzero, \Dplus$ and $\Dstar$ $\Rppb$. Within uncertainties, the measured D-meson $\Rppb$ is compatible with unity, with hints of small ($<10$--$20\%$) CNM effects for $\pt < 2~\gevc$. This confirms the large suppression of D mesons in Pb--Pb collisions is indeed a final state effect due to the interaction with the hot, deconfined medium. The D-meson $\Rppb$ is compared to models - in the left panel models are shown which include only CNM effects, including NLO pQCD predictions \cite{MANGANO1992295} with EPS09 parameterisation of the nuclear parton distribution function (nPDF) \cite{Eskola:2009uj}, a Leading Order pQCD calculation including shadowing, $k_T$ broadening and CNM energy loss \cite{Sharma:2009hn}, a model based on the Colour Glass Condensate \cite{Fujii:2013yja}, and a higher-twist calculation based on incoherent multiple scattering \cite{Kang:2014hha}. All models are consistent with the data except that based on incoherent multiple scattering, which is disfavoured by the data at low $\pt$.
	The right panel shows a comparison with models which assume a Quark--Gluon Plasma is formed - the DUKE model \cite{Xu:2015iha} includes both collisional and radiative processes, and the POWLANG model \cite{Beraudo:2015wsd} includes only collisional processes, considering two different transport coefficients.
	While the data are consistent with the models within uncertainties, the shape of the $\Rppb$ slightly disfavours these models. The comparison with models shows the motivation for a precise measurement at low $\pt$, since this is where the models deviate from each other.
	
\begin{figure}[hbt]
		\centering
		\includegraphics[width=.47\textwidth]{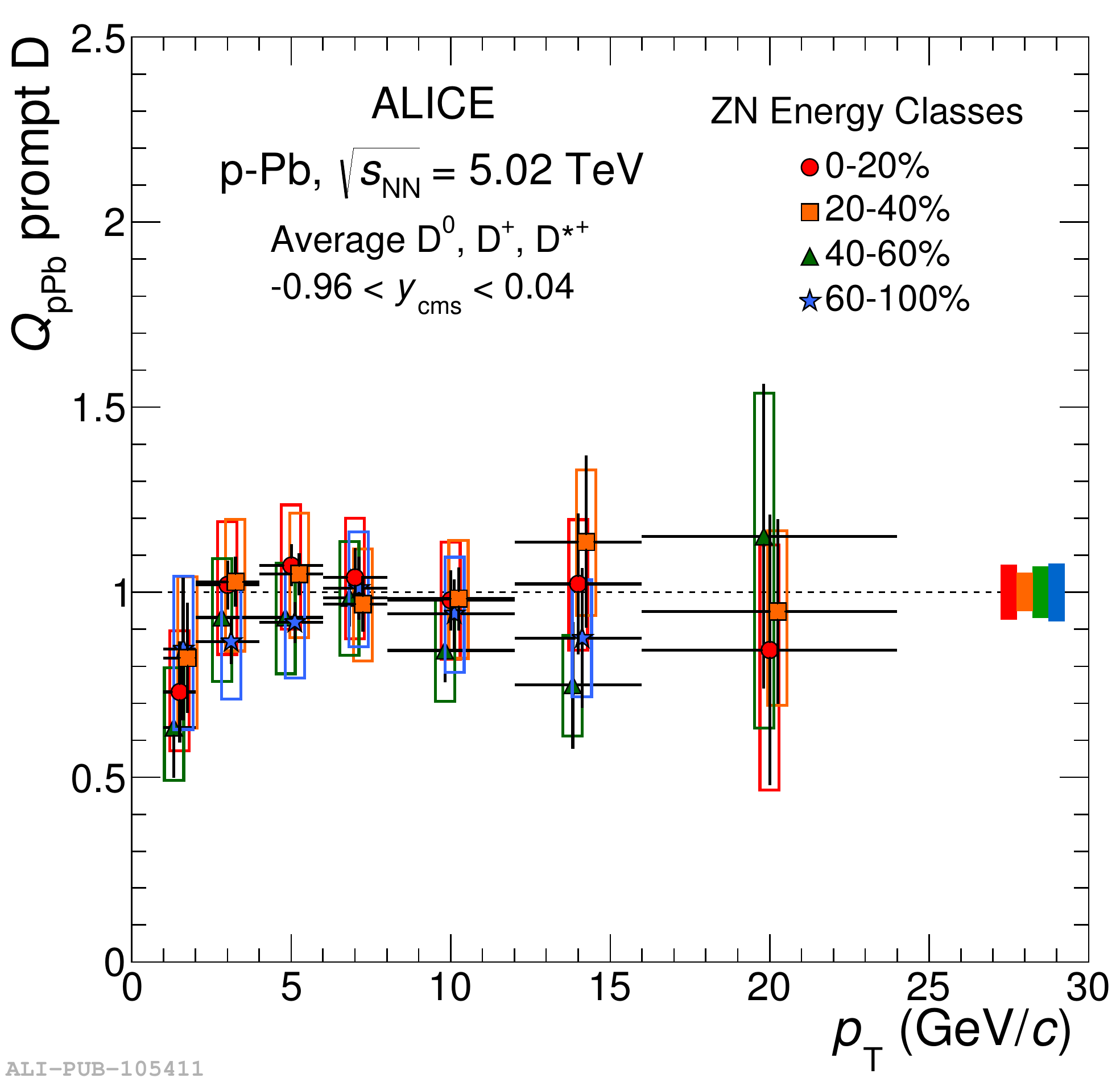}
		\caption[$\Qppb$ of D mesons measured by ALICE]{The averaged D-meson centrality dependent nuclear modification factor $\Qppb$ in the centrality classes 0-20\%, 20-40\%, 40-60\% and 60-100\%, using the ZNA centrality estimator \cite{Adam:2016mkz}. }  
		\label{fig:D0QpPb-ALICE}
	\end{figure}

	The multiplicity dependence of the nuclear modification factor has also been studied for D mesons by measuring the `centrality-dependent' nuclear modification factor $\Qppb$ given by
	
	\begin{equation}
	\Qppb = \frac{1}{\langle T_\mathrm{pPb}^{cent}\rangle} 
	\frac{\mathrm{d}N_\mathrm{pPb}^{cent} / \mathrm{d}\pt}
	{\mathrm{d}\sigma_\mathrm{pp} / \mathrm{d}\pt}
	\end{equation}
	
	where $\mathrm{d}N_\mathrm{pPb}^{cent} / \dpt$ is the D-meson yield in p--Pb collisions in a given centrality class, $\mathrm{d}\sigma_\mathrm{pp} / \dpt$ is the D-meson cross section in pp collisions and $\langle T_{\mathrm{pPb}}^{cent} \rangle$ is the average nuclear overlap function in a given centrality class, equal to the number of binary nucleon--nucleon collisions divided by the inelastic nucleon--nucleon cross section.
	The centrality in p--Pb collisions can be estimated using different event-activity estimators, which can generate dynamical biases if based on multiplicity measurements (see e.g. \cite{Adam:2014qja}). Figure \ref{fig:D0QpPb-ALICE} shows the $\Qppb$ measured in p--Pb collisions at $\sqrtSnn = 5.02~\tev$ \cite{Adam:2016mkz} where the centrality is estimated using the energy deposited in the Zero Degree Neutron Calorimeter (ZNA) by slow nucleons produced in the collision by nuclear de-excitation processes, or knocked out by wounded nucleons. This centrality estimator does not rely on a multiplicity measurement and thus is expected to generate the least bias. 
	The $\Qppb$ in all centrality classes are seen to be consistent with each other and with unity, suggesting there is no centrality dependence in the nuclear modification of D mesons within the experimental uncertainties.

\begin{figure}[hbt]
		\centering
		\includegraphics[width=.42\textwidth]{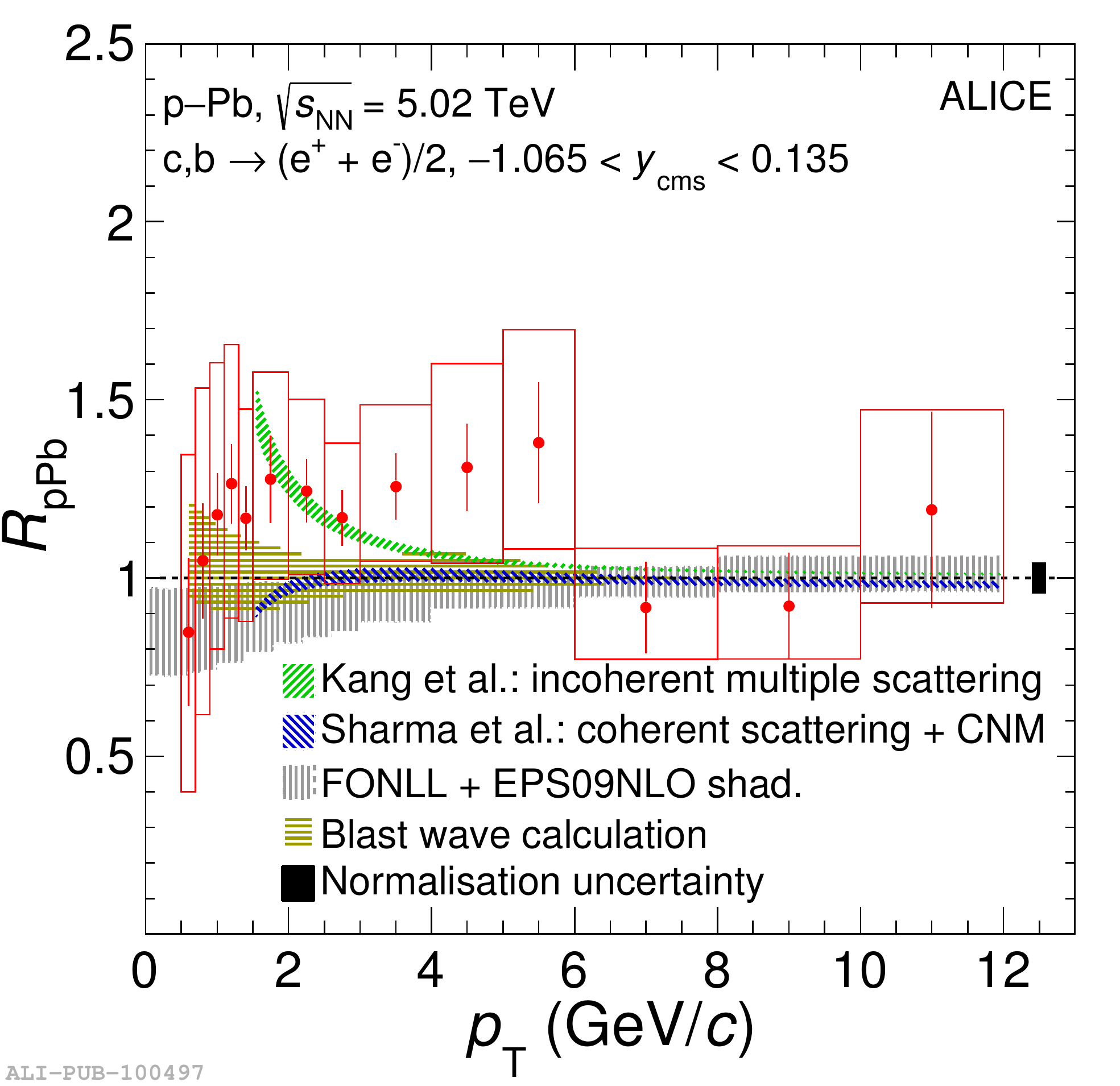}
		\includegraphics[width=.42\textwidth]{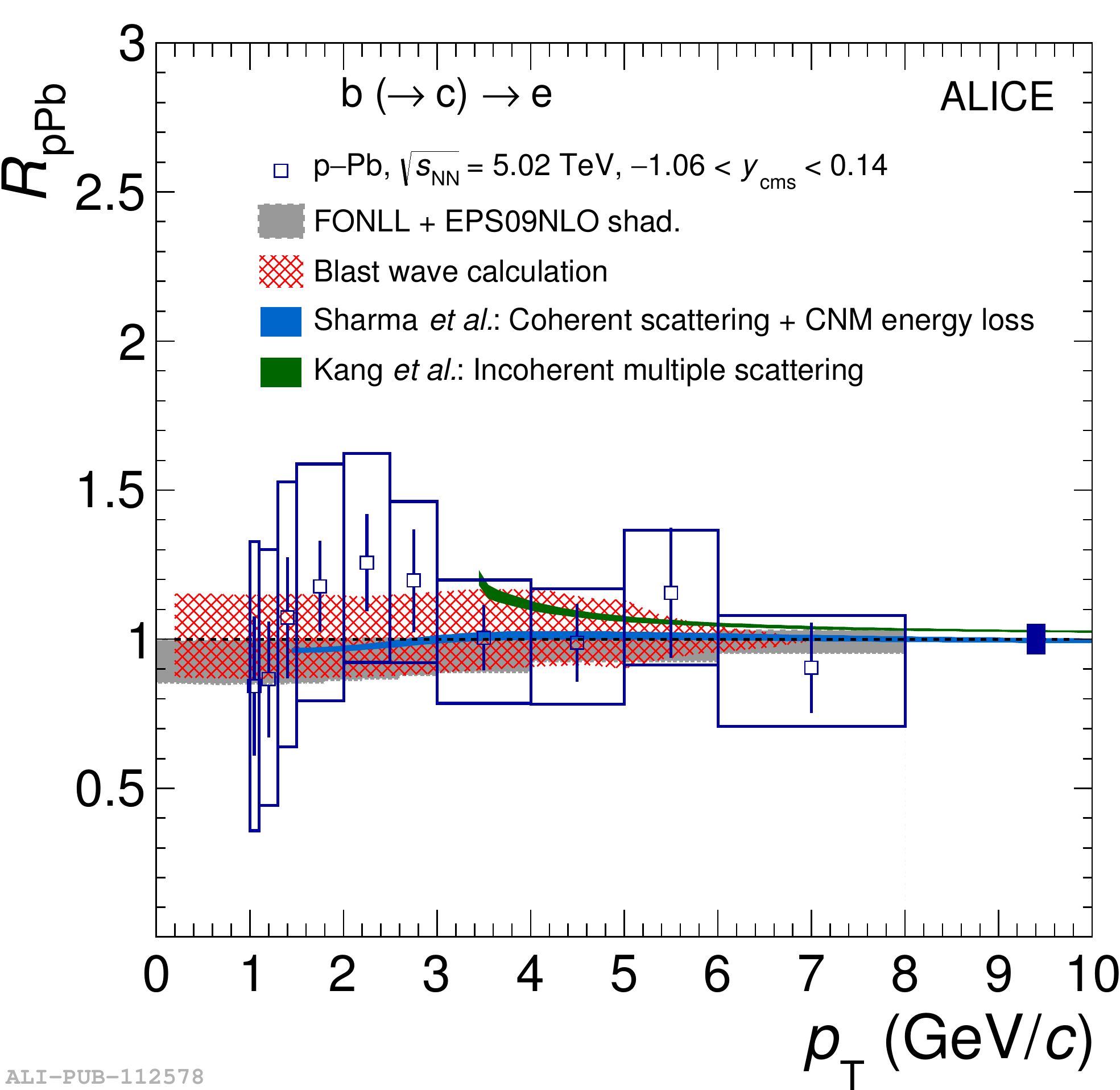}
		\includegraphics[width=.5\textwidth]{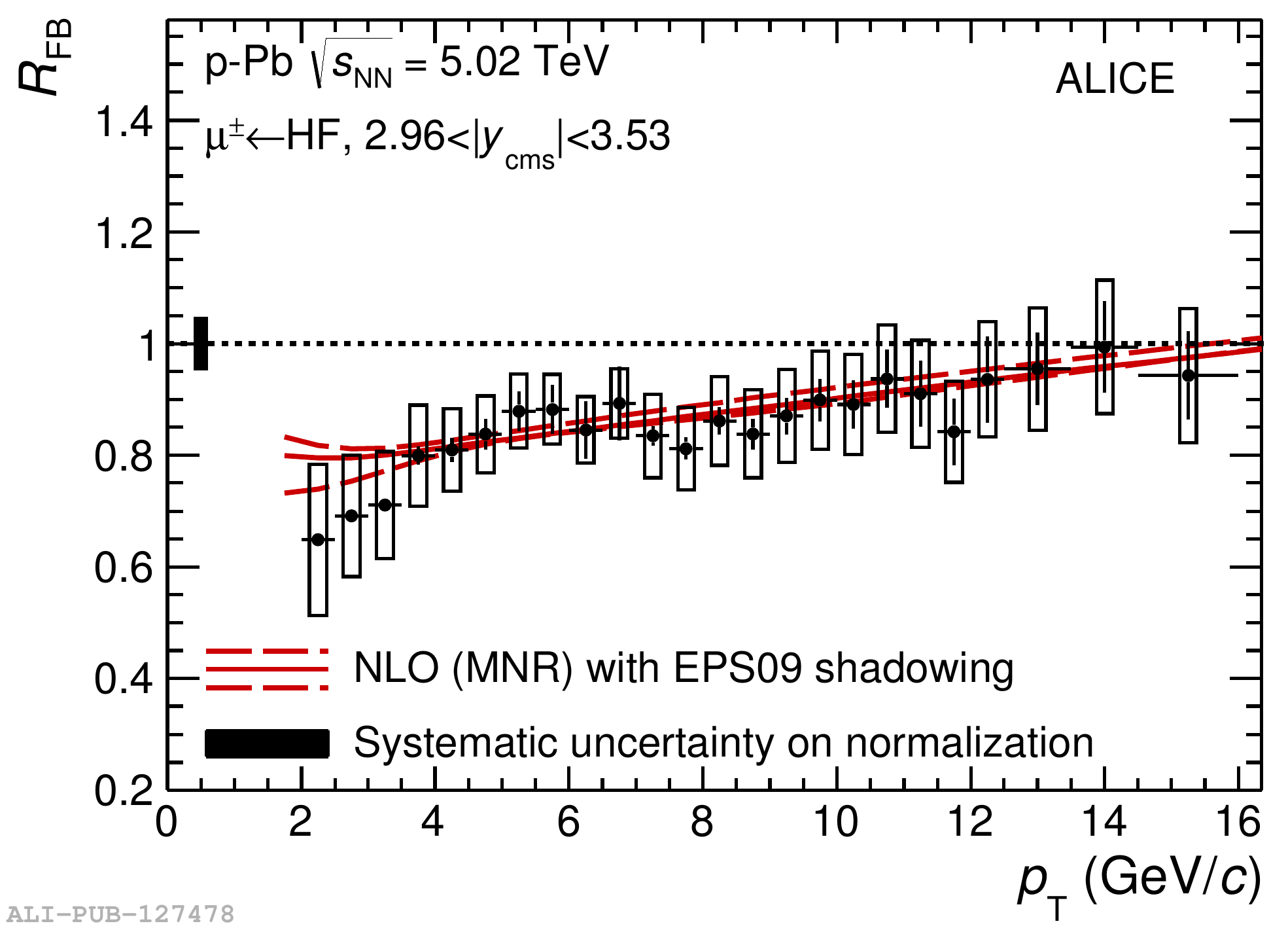}
		\caption[]{Heavy-flavour lepton measurements in p--Pb collisions. Top panels: the $\Rppb$ of electrons from heavy-flavour hadron decays (left) \cite{Adam:2015qda} and the $\Rppb$ of electrons from beauty-hadron decays (right) \cite{Adam:2016wyz} compared to model predictions \cite{Cacciari:1998it,Eskola:2009uj,Sharma:2009hn,Kang:2014hha,Sickles:2013yna}. Bottom panel: The forward-to-backward ratio of muons from heavy-flavour hadron decays \cite{Acharya:2017hdv}, compared to model predictions \cite{MANGANO1992295,Eskola:2009uj}.}  
		\label{fig:lepton-pPb}
	\end{figure}

	The nuclear modification of electrons from heavy-flavour decays has been measured in p--Pb collisions, for electrons from heavy-flavour (charm and beauty) hadron decays \cite{Adam:2015qda} and electrons from beauty decays \cite{Adam:2016wyz}. The top panels of figure \ref{fig:lepton-pPb} show the respective $\Rppb$ measurements in comparison with model predictions, including FONLL \cite{Cacciari:1998it} with EPS09 nPDF parameterisation \cite{Eskola:2009uj}, the same models as shown for D mesons based on either coherent or incoherent scattering \cite{Sharma:2009hn,Kang:2014hha}, and a blast wave calculation to understand the effect of the possible hydrodynamical expansion of the system on charm and beauty quarks (as done in \cite{Sickles:2013yna}), all of which describe the data within the experimental uncertainties.
	Muons from heavy-flavour hadron decays have also been measured in p--Pb collisions \cite{Acharya:2017hdv} where the beam direction is switched, allowing for measurements at both forward (p-beam travelling towards the muon spectrometer) and backward (Pb-beam travelling towards the muon spectrometer) rapidities. The $\Rppb$ has been measured for both configurations, and the forward-backward ratio, defined as the ratio of the cross section of muons from heavy-flavour hadron decays at forward rapidity to that at backward rapidity in a rapidity interval symmetric
	with respect to $y_{cms} = 0$, is measured, as shown in the bottom panel of figure \ref{fig:lepton-pPb}. The forward-to-backward ratio is measured to be below unity at low $\pt$, and is described by NLO pQCD predictions \cite{MANGANO1992295} with the EPS09 nPDF parameterisation \cite{Eskola:2009uj}.
	
	\section{Summary and outlook}
	\label{sec:summary}
	
	The measurement of heavy quarks produced in pp and p--Pb collisions with ALICE in Run 1 of the LHC has enabled tests of pQCD predictions and quark fragmentation properties, and has provided important constraints into the study of cold nuclear matter effects present in p--Pb collisions. 
	
	Higher statistics datasets collected during the recent pp and p--Pb runs at the LHC during Run 2 will allow for higher-precision measurements of heavy-flavour production. Finally, the upgrade program for Run 3 includes a full ITS upgrade \cite{ITS-upgrade}, a Muon Forward Tracker (MFT) and continuous TPC readout which will allow for a large precision and statistical improvement in all these measurements, and will open up the possibility to study new heavy-flavour observables (e.g. B mesons, $\mathrm{\Lambda_{c}}$ baryons) in Pb--Pb collisions.


	\bibliographystyle{JHEP}
	\bibliography{my-bib-database}
	
\end{document}